\documentclass[%
 twocolumn,
 amsmath,amssymb,
 aps, 
]{revtex4-2}

\usepackage{graphicx}% Include figure files
\usepackage{dcolumn}% Align table columns on decimal point
\usepackage{physics}% bracket
\usepackage{booktabs}
\usepackage{float}
\usepackage[usenames, dvipsnames]{xcolor}
\usepackage{amsmath}
\usepackage{newtxtext, newtxmath}
\usepackage{multirow}
\DeclareSymbolFont{CMlargesymbols}{OMX}{cmex}{m}{n}
\let\sumop\relax\let\prodop\relax
\DeclareMathSymbol{\sumop}{\mathop}{CMlargesymbols}{"50}
\DeclareMathSymbol{\prodop}{\mathop}{CMlargesymbols}{"51}

\SetSymbolFont{operators}{normal}{OT1}{ntxtlf}{m}{n}
\SetSymbolFont{operators}{bold}{OT1}{ntxtlf}{b}{n}

\renewcommand{\vec}[1]{\boldsymbol{#1}}

\newcommand{\ee}{\mathrm{e}}
\newcommand{\ii}{\mathrm{i}}

\DeclareMathOperator*{\QFT}{QFT}

\renewcommand{\ket}[1]{| #1 \rangle}
\renewcommand{\bra}[1]{\langle #1 |}

\begin{document}

% \preprint{APS/123-QED}

%\linenumbers

\title{Quantum computing of the nonlinear Schrödinger equation via measurement-induced potential reconstruction}
% Numerical simulation of the nonlinear Schrödinger equation using the quantum split-step Fourier method

\author{Kaiwen Weng}
\affiliation{
 Shanghai Institute of Applied Mathematics and Mechanics, School of Mechanics and Engineering Science, Shanghai University, Shanghai 200444, PR China\\
 The State Key Laboratory of Nonlinear Mechanics, Institute of Mechanics, Chinese Academy of Sciences, Beijing 100190, PR China
 }

\author{Zhaoyuan Meng}
\email{mengzhaoyuan@imech.ac.cn}
\affiliation{
 The State Key Laboratory of Nonlinear Mechanics, Institute of Mechanics, Chinese Academy of Sciences, Beijing 100190, PR China
}

% \author{Zixuan Yang}
% \email{yangzx@imech.ac.cn}
% \affiliation{
%  The State Key Laboratory of Nonlinear Mechanics, Institute of Mechanics, Chinese Academy of Sciences, Beijing 100190, PR China
% }

\author{Guohui Hu}
\email{ghhu@staff.shu.edu.cn}
\affiliation{
Shanghai Institute of Applied Mathematics and Mechanics, School of Mechanics and Engineering Science, Shanghai University, Shanghai 200444, PR China
}

\date{\today}

\begin{abstract}
The nonlinear Schrödinger equation (NLSE) is a fundamental model that describes diverse complex phenomena in nature.
However, simulating the NLSE on a quantum computer is inherently challenging due to the presence of the nonlinear term.
We propose a hybrid quantum-classical framework for simulating the NLSE based on the split-step Fourier method.
During the linear propagation step, we apply the kinetic evolution operator to generate an intermediate quantum state.
Subsequently, the Hadamard test is employed to measure the Fourier components of low-wavenumber modes, enabling the efficient reconstruction of nonlinear potentials.
The phase transformation corresponding to the reconstructed potential is then implemented via a quantum circuit using the phase kickback technique.
To validate the efficacy of the proposed algorithm, we numerically simulate the evolution of a Gaussian wave packet, a soliton wave, snake instability, and the wake flow past a cylinder.
The simulation results demonstrate excellent agreement with the corresponding classical solutions.
This work provide a concrete basis for analyzing accuracy-cost trade-offs in quantum-classical simulations of nonlinear dispersive wave dynamics.
\end{abstract}

\maketitle

\section{\label{sec:1}Introduction}

The nonlinear Schrödinger equation (NLSE) constitutes a canonical model for quantum phenomena, principally governing the mean-field dynamics of Bose-Einstein condensates (BEC)~\cite{dalfovo1999theory}.
Extending beyond the quantum regime, the NLSE arises in macroscopic systems and plays a central role in the description of weakly nonlinear, narrow-banded wave trains within fluid mechanics.
This equation captures the interplay between dispersion and nonlinearity that underlies modulational instability~\cite{benjamin1967disintegration,zakharov1968stability,hasimoto1972nonlinear}, self-focusing~\cite{shabat1972exact}, and defocusing~\cite{tsuzuki1971nonlinear}, a versatility that has been substantiated by extensive theoretical, numerical, and experimental studies~\cite{yuen1975nonlinear,lake1977nonlinear,osborne2000nonlinear,dysthe1979note,Hardin1973ApplicationOT}.

The multidimensional NLSE serves as a fundamental framework for describing wave-packet dynamics in nonlinear media.
Unlike its one-dimensional counterpart, the NLSE in two or higher dimensions is generally non-integrable, giving rise to complex phenomena such as wave collapse and quantum turbulence.
Classical numerical simulations of such systems are hindered by severe scalability constraints.
Specifically, the computational cost scales polynomially with the discretization grid size and increases exponentially with the spatial dimension due to the curse of dimensionality.
In this context, quantum computing emerges as a transformative paradigm with significant potential for advancing this field.
By leveraging Hamiltonian simulation algorithms, quantum platforms enable the encoding of dynamics with logarithmic resource scaling, thereby offering a powerful route to investigate high-dimensional non-integrable physics beyond the reach of classical methods.

Recent advancements in quantum hardware and algorithms have stimulated significant interest in quantum approaches for the numerical solution of partial differential equations (PDEs) and high-dimensional dynamical systems.
In the context of fluid mechanics and related continuum problems, existing methodologies can be broadly categorized into two distinct paradigms, distinguished by their treatment of temporal evolution and problem encoding.

The first research avenue centers on direct dynamical encoding, where the primary objective is to map physical evolution onto unitary quantum circuits, thereby mitigating the measurement and reinitialization overheads inherent to classical time-marching schemes.
The Schrödingerization framework~\cite{jin_quantum_maxwell_2024,jin_quantum_fokker_2024,jin_quantum_partial_2024,jin_schrodingerization_2025,jin_quantum_heat_2025} provides a unified construction for general linear PDEs by embedding non-unitary operators into higher-dimensional unitary dynamics through ancillary qubits.
In a similar vein, the linear combination of Hamiltonians offers an alternative route that maps general linear non-unitary dynamics to a higher-dimensional unitary representation~\cite{An2023_Linear, An2023_Quantum, Meng2025_Toward, Low2025_Optimal}.
Besides, quantum spin representations of the Navier-Stokes equations provide a continuous framework for mapping fluid dynamics directly onto the evolution of a quantum system~\cite{meng2023quantum,meng2024lagrangian,meng2024quantum,meng2024simulating}.
Moreover, quantum lattice Boltzmann methods are consistent with this Hamiltonian-compatible framework via the mapping of their streaming and collision processes onto unitary operators~\cite{zamora2025efficient,singh2025emergence,kocherla2024fully,wang2025quantum}.

The second major paradigm involves solver-based hybrid quantum-classical methods, wherein the temporal integration and nonlinear treatments are retained within a classical framework, while quantum routines serve as subroutines to accelerate specific linear-algebraic tasks~\cite{chen2022quantum,lapworth2022hybrid,liu2023quantum,bharadwaj2023hybrid,ye2024hybrid,chen2024enabling,bharadwaj2024compact}.
This approach typically addresses the computationally intensive inversion of stiffness matrices or the solution of pressure Poisson equations through quantum linear algebra algorithms, such as the Harrow-Hassidim-Lloyd algorithm~\cite{harrow2009quantum}, quantum singular value transformation~\cite{gilyen2019quantum}, and variational quantum linear solvers~\cite{xu2021variational}.
In contrast to Hamiltonian simulation, these hybrid schemes prioritize the exploitation of quantum advantages for discrete matrix operations over the direct encoding of continuous physical dynamics.

Although quantum algorithms for linear algebra and linear dynamics are well developed, their extension to nonlinear problems constitutes a fundamental challenge.
This difficulty arises because quantum mechanics is intrinsically governed by linear unitary evolution.
To reconcile this linearity with nonlinear dynamics, two principal strategies have emerged.
One strategy embeds the nonlinear system within a higher-dimensional linear Hilbert space~\cite{Joseph2020_Koopman, Liu2021_Efficient, Jin2023_Time, Succi2024_Ensemble, Tennie2025_Integration}, which enables unitary simulation at the expense of increased qubit resources and complexity.
Alternatively, nonlinearity can be treated through hybrid protocols that rely on measurement, feedback, and renormalization, utilizing the collapse of the wave function to extract nonlinear information.

Within this landscape, the NLSE presents a uniquely favorable case.
Its nonlinearity depends explicitly on the probability density, a quantity that is naturally accessible through direct quantum measurement.
Crucially, the NLSE governs a Hamiltonian system with inherent unitary evolution.
These characteristics distinguish the NLSE from general dissipative nonlinear PDEs, rendering it exceptionally well-suited for quantum computing.
Despite these intrinsic advantages, quantum algorithmic research targeting the NLSE remains remarkably limited.
Current efforts include a variational-quantum-eigensolver-based approach to approximate the nonlinear step within a split-step Fourier method (SSFM)~\cite{kocher2025numerical} and quantum-inspired tensor network methods~\cite{connor2025tensor}.

Here, we present a hybrid quantum-classical split-step spectral framework for simulating the NLSE.
The method combines quantum Fourier transform (QFT) for linear propagation with measurement-assisted local-phase updates for the nonlinear regime, thereby maintaining globally unitary evolution through sequences of quantum gates.
We devise and analyze a spectral filtering strategy that retains dominant modes while reducing effective dimensionality, enabling stable temporal integration under restricted qubit and measurement constraints.
Computational complexity analysis and comparisons with classical algorithms are provided.
Furthermore, we validate the efficacy of the present framework by numerically simulating 1D solitons, 2D Gaussian wave packets, 2D snake instability, and 2D wake flow past a cylinder.

The paper is organized as follows.
In Sec.~\ref{sec:2}, we introduce the NLSE model and the classical SSFM.
Section~\ref{sec:3} details the quantum split-step Fourier method (QSSFM) framework, specifically addressing the implementation of filtering and the Hadamard test.
Numerical validations by quantum simulation on a classical computer are presented in Sec.~\ref{sec:4}.
Conclusions and discussion are drawn in Sec.~\ref{sec:5}.

\begin{figure*}[t!]
    \centering
    \includegraphics{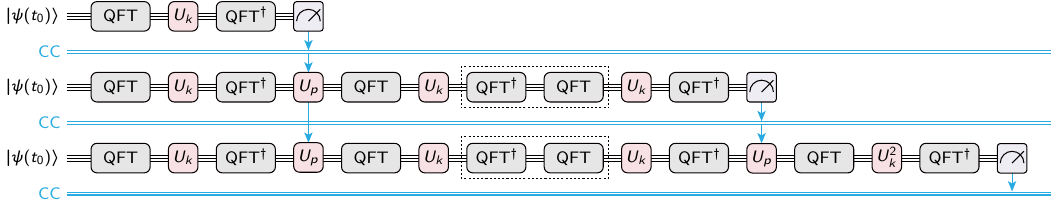}
    \caption{Schematic of the first three QSSFM time-step iterations.
    Blue lines denote the classical computer (CC).
    The operators $U_k$ and $U_p$ implement linear and nonlinear propagation, respectively.
    The operators enclosed in the dashed box represent computational processes that can be canceled.
    The measurement block extracts selected Fourier modes of the wave function.
    Following a Fourier transform on the CC, the approximate nonlinear potential is utilized to reconstruct the nonlinear operator $U_p$ for the subsequent quantum evolution step.}
    \label{fig:qcircuit}
\end{figure*}

\begin{figure}[t!]
    \centering
    \includegraphics{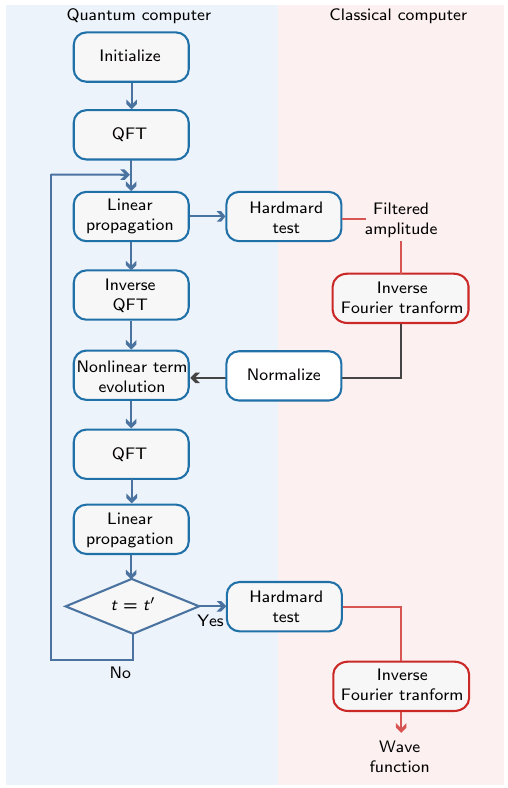}
    \caption{Workflow of the filtered-QSSFM.
    The blue and pink regions represent the quantum and classical computation components, respectively.
    Rectangular boxes indicate computational steps, while the diamond-shaped box checks whether the current time $t$ equals the final time $t'$.
    Normalization of the values following the inverse Fourier transform enhances the simulation accuracy.}
    \label{fig:workflow}
\end{figure}

\begin{figure}[t!]
    \centering
    \includegraphics{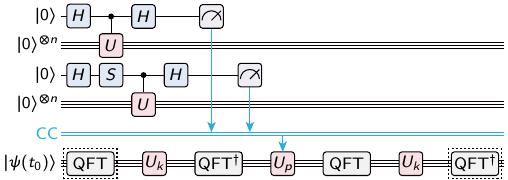}
    \caption{Quantum circuit of the filtered-QSSFM incorporating the Hadamard test.
    Here, $H$ and $S$ denote the Hadamard gate and the phase gate, respectively, and $U$ is defined as $U = U_{\ell}^\dagger U_{\psi}$.
    The two Hadamard tests, implemented with and without the phase gate, correspond to measuring the real and imaginary parts of the retained mode.}
    \label{fig:Hardmard_test}
\end{figure}

\section{\label{sec:2}Nonlinear Schrödinger equation and classical split-step Fourier method}
\subsection{\label{subsec:2.1}Nonlinear Schrödinger equation}

The standard form of the NLSE is
\begin{eqnarray}\label{equ:1}
    \ii \hbar \frac{\partial\psi(\vec{x},t)}{\partial t} = \left( -\frac{\hbar}{2m}\nabla^2 +g|\psi(\vec{x},t)|^2 + V(\vec{x},t)\right) \psi(\vec{x},t),
\end{eqnarray}
where $\psi(\vec{x},t)$ the wave function, $V(\vec{x},t)$ the external potential, $g\in\mathbb{R}$ the coupling coefficient, $\hbar$ the reduced Planck constant, $m$ the particle mass, and $\ii=\sqrt{-1}$. 
For simplicity, we adopt natural units by setting $\hbar = m = 1$. 

By defining the linear operator $\mathcal{H}_k = -\frac{1}{2}\vec{\nabla}^2$ and the nonlinear operator $\mathcal{H}_p = g|\psi(\vec{x},t)|^2 + V(\vec{x},t)$, Eq.~(\ref{equ:1}) is recast as
\begin{equation}
\ii\frac{\partial\psi(\vec{x},t)}{\partial t} = \left( \mathcal{H}_k + \mathcal{H}_p \right) \psi(\vec{x},t).
\label{equ:2}
\end{equation}
The temporal evolution of the wave function $\psi(\vec{x},t)$ is simulated using a unitary time-evolution operator $U(t,t+\tau)$, which propagates the solution over a small time step $\tau$ according to
\begin{equation}
\psi(\vec{x},t_j + \tau) = U(t,t+\tau)\psi(\vec{x},t_j),
\end{equation}
where the exact time-evolution operator for the NLSE is given by
\begin{equation}
U(t_j,t_j+\tau) = \ee^{- \ii \tau (\mathcal{H}_k+\mathcal{H}_p)}.
\end{equation}

\subsection{\label{subsec:2.2}Classical split-step Fourier method}
The computational efficiency of the SSFM derives from an operator-splitting scheme that effectively decouples the components of the Hamiltonian.
Based on the symmetric Strang splitting \cite{strang1968construction}, the time-evolution operator is approximated as
\begin{equation}\label{equ:5}
    \begin{aligned}
        \psi(\vec{x},t_j +\tau) & = U(t_j,t_j+\tau)\psi(\vec{x},t_j) 
        \\
        & = \ee^{-\ii\tau(\mathcal{H}_k+\mathcal{H}_p)}\psi(\vec{x},t_j) 
        \\
        & \approx \ee^{-\ii\frac{\tau}{2}\mathcal{H}_k}\ee^{-\ii\tau\mathcal{H}_p}\ee^{-\ii\frac{\tau}{2}\mathcal{H}_k}\psi(\vec{x},t_j).
    \end{aligned}
\end{equation}
This splitting is exact provided that the commutator $\left[\mathcal{H}_k, \mathcal{H}_p\right]$ vanishes. Otherwise, the method is second-order accurate with a local error scaling as $\mathcal{O}(\tau^3)$.
Such a decomposition facilitates the exact integration of the subproblem driven by $\mathcal{H}_k$ via spectral methods, utilizing the Fourier transform property
\begin{equation}
    \mathcal{H}_k = -\frac{1}{2}\nabla^2 \xrightarrow{\mathcal{F}}\frac{1}{2}\vec{k}^2,
\end{equation}
where $\vec{k}$ denotes the wave vector and $\mathcal{F}$ represents the Fourier transform.
Consequently, the evolution operator in the spectral space takes the form
\begin{equation}\label{equ:7}
    U(t_j, t_j+\tau)
    = \mathcal{F}^{-1} \ee^{-\ii\frac{\tau}{4}k^2}\mathcal{F} \ee^{-\ii\tau\mathcal{H}_p} \mathcal{F}^{-1} \ee^{-\ii\frac{\tau}{4}k^2}\mathcal{F},
\end{equation}
where $\mathcal{F}^{-1}$ denotes the inverse Fourier transform.

\section{\label{sec:3}Quantum Split-Step Fourier Method}
The QSSFM constitutes a quantum-mechanical adaptation of the classical SSFM, in which the conventional operators are replaced by their quantum counterparts according to $\mathcal{H}_k \rightarrow \hat{\mathcal{H}_k}$, $\mathcal{H}_p \rightarrow \hat{\mathcal{H}_p}$, and $k \rightarrow \hat{k}$.
This transformation permits the simulation of the NLSE dynamics on a quantum processor while maintaining the essential structural characteristics of the classical algorithm.

The discretized wave function samples $\psi(t_j)$ are encoded into the amplitudes of an $n$-qubit state $\ket{\psi(t_j)}$ according to
\begin{align}
    \ket{\psi(t_j)}=  \sum_{xyz}\frac{\psi(x,y,z,t_j)}{\|\psi(x,y,z,t_j)\|_2^2}\ket{xyz},
\end{align}
where the indices $x,y,z$ represent the discrete spatial coordinates.

\subsection{\label{subsec:3.1}Linear propagation}
% According to Eq.~(\ref{equ:7}), the linear term becomes
% \begin{equation}
%     \ee^{-\ii\frac{\tau}{2}\hat{\mathcal{H}}_k}\ket{\psi(t_j)} 
%     = {\QFT}^\dagger \ee^{-\ii\frac{\tau}{4}\hat{k}^2}\QFT\ket{\psi(t_j)}
%     \label{equ:9}
% \end{equation}
% where $\QFT^\dagger$ denotes the inverse quantum Fourier transform.

% In principle, the diagonal phase operator $e^{-i\frac{\tau}{4}\hat{k}^2}$ can be decomposed into elementary gates~\cite{bergholm2005quantum}. In our setting, its diagonal entries are fixed for a given grid, so we precompile it into a single unitary gate $U_k$ to reduce compilation overhead.

In the linear propagation step, the quantum state evolves according to
\begin{equation}
    \ee^{-\ii\frac{\tau}{2}\hat{\mathcal{H}}_k}\ket{\psi(t_j)} = {\QFT}^\dagger U_k \QFT\ket{\psi(t_j)},
    \label{equ:13}
\end{equation}
where the diagonal operator is defined as
\begin{equation}
    U_k  = \sum_{\ell=0}^{N-1} \ee^{-\ii \tau k_\ell^2/4} \ket{\ell}\bra{\ell}
\end{equation}
with $N=2^n$.
By utilizing the decomposition~\cite{meng2024simulating}
\begin{equation}
    \sum_{\ell=0}^{N-1} k_\ell \ket{\ell}\bra{\ell}
    = -\frac12\bigg(I_{2^n} + \sum_{j=1}^n 2^{n-j}Z_j \bigg) + 2^{n-1}Z_1,
\end{equation}
the unitary operator $U_k$ is implemented with a complexity of $\mathcal{O}(n^2)$ basic gates.
Here, $Z_j=I_{2^j-1} \otimes Z \otimes I_{2^{n-j}}$ denotes the phase-shift gate acting on the $(n-j+1)$-th qubit, and $Z$ represents the Pauli-$Z$ gate.

\subsection{\label{subsection:3.2}Nonlinear potential}

The nonlinear evolution in the QSSFM is implemented via a diagonal phase gate $U_p$, which encodes the interaction potential based on the probability density of the wave function. Since the linear kinetic term is decoupled from the nonlinear interaction, the operator is explicitly defined as
\begin{equation}
    U_p := \ee^{-\ii\tau\hat{\mathcal{H}}_p}= \sum_{\ell=0}^{N-1} \ee^{-\ii\tau \left[ |\psi(x_\ell, t_j)|^2+V(\vec{x},t_j) \right]} \ket{x_\ell}\bra{x_\ell},
\end{equation}
where $\ket{x_\ell}$ denotes the position basis. As illustrated in Fig.~\ref{fig:qcircuit}, we optimize the circuit depth by merging the last linear step of the previous iteration with the first linear step of the current iteration (dashed box), thereby reducing gate complexity while maintaining algorithmic accuracy.

However, a direct implementation of this procedure faces a fundamental bottleneck: measuring the probability density to evaluate $U_p$ destroys quantum coherence, interrupting the continuous unitary evolution. To proceed with time-stepping, the state must be reconstructed, and relying on full quantum state tomography incurs a prohibitive overhead in both measurement counts and classical post-processing. Consequently, it is essential to develop a strategy that minimizes measurement costs while preserving the fidelity of the simulation relative to the classical reference.

To address this limitation, we propose the filtered-QSSFM, which utilizes the spectral representation to reduce the measurement overhead.
By alternating between position and spectral spaces via Fourier transforms, a subset of spectral components is selectively measured, specifically by discarding high-wavenumber modes with negligible amplitudes.
This spectral filtering serves as a control mechanism, allowing for a trade-off between simulation accuracy and measurement complexity, analogous to high-frequency truncation in classical numerical schemes.

We employ the Hadamard test to extract the complex spectral coefficients of the selected modes without collapsing the main simulation register.
As depicted in Fig.~\ref{fig:Hardmard_test}, the circuit architecture decouples the unitary evolution from the readout process, such that the main evolution circuit propagates the state forward in time while measurements are performed in auxiliary readout circuits.
The Hadamard test utilizes the unitary operator $U=U_{\ell}^\dagger U_\psi$, satisfying
\begin{equation}
    \bra{0}^{\otimes n} U_{\ell}^\dagger U_\psi\ket{0}^{\otimes n} = \braket{\ell}{\psi},
\end{equation}
where $U_{\psi}$ denotes the circuit for preparing $\ket{\psi}$ and $U_{\ell}$ prepares the target computational basis state $\ket{\ell}$.
For a given $\ell$, $U_\ell$ is implemented as a sequence of $X$ gates.
To obtain the complex coefficient for each of the $M$ retained modes, we execute two distinct circuits to measure the real component $\mathrm{Re}(\braket{\ell}{\psi})$ and the imaginary component $\mathrm{Im}(\braket{\ell}{\psi})$, respectively.
These measurements are collected via the classical information channel (see Fig.~\ref{fig:Hardmard_test}) and repeated to ensure statistical convergence, all while the primary quantum state remains unperturbed.

The overall hybrid quantum-classical workflow is illustrated in Fig.~\ref{fig:workflow}.
The spectral coefficients extracted via the Hadamard test are processed on a classical computer using an inverse Fourier transform to reconstruct the wave function in position space.
At this stage, the reconstructed amplitudes may optionally be renormalized prior to re-encoding; the impact of this renormalization on numerical stability is examined in Sec.~\ref{sec:4}.
This reconstructed information is subsequently utilized to construct the nonlinear operator $U_p$.
By assigning the measured values to their corresponding positions and setting unmeasured diagonal entries to zero, $U_p$ is formulated as a sparse, diagonally dominant matrix, which facilitates efficient implementation within the quantum circuit~\cite{welch2014efficient}.

\subsection{\label{subsec:3.5}Complexity analysis}
\begin{table*}[t!]
    \caption{Summary of the computational complexity for the classical SSFM, the standard QSSFM (assuming full tomography), and the proposed Filtered-QSSFM. Here, $N_t$ denotes the total number of time steps, $N$ the grid size, $M$ the number of retained modes, and $\varepsilon$ the target precision for the nonlinear potential reconstruction.} 
    \label{tab:complexity}
    \begin{ruledtabular}
        % \begin{tabular}{ccccccc}
        %                  & Fourier     & Linear & Inverse Fourier  & Nonlinear   & Measurement  & Total \\ %换行
        %                  &transform    & propagation &  transform       & evolution  &             & step  \\ %换行
        %     \midrule %[2pt] 
        %     SSFM   & $\mathcal{O}(N\log N)$ & $\mathcal{O}(N)$ & $\mathcal{O}(N\log N)$ & $\mathcal{O}(N)$ & -- & $\mathcal{O}(N\log N)$\\
        %     QSSFM        & $\mathcal{O}(\log^2 N)$     & $\mathcal{O}(\log^2 N)$ & $\mathcal{O}(\log^2 N)$     & $\mathcal{O}(N)$ & $\mathcal{O}(N^2)$& $\mathcal{O}(N^2)$    \\
        %     Filtered-QSSFM & $\mathcal{O}(M\log M)$     & $\mathcal{O}(\log^2 N)$ & $\mathcal{O}(\log^2 N)$   & $\mathcal{O}(M)$ & $\mathcal{O}(M \log^2 N)$& $\mathcal{O}(M \log^2 N)$    \\
        %     % VQA for NLSE & /                      & /                & /                      & /               &/ & $\mathcal{O}(\log^2 N)$     \\
        % \end{tabular}
% \usepackage{multirow}
    \renewcommand{\arraystretch}{1.3}
	\begin{tabular}{ccccc}
		Method         & Spatial scaling                & Temporal scaling                     & Sampling cost                         & Total complexity                                     \\ 
		               & (per step)                     & (State prep)                         & (per step)                            &                                                      \\ \hline
		SSFM           & $\mathcal{O}(N \log N)$                  & Linear ($N_t$)                       & Deterministic                         & $\mathcal{O}(N_t N\log N)$                                     \\
		Full QSSFM     & \multirow{2}{*}{$\mathcal{O}(\log^2 N)$} & \multirow{2}{*}{Quadratic ($N_t^2$)} & \multirow{2}{*}{$\mathcal{O}(N/\varepsilon^2)$} & \multirow{2}{*}{$\mathcal{O}(N_t^2 N\log^2 N /\varepsilon^2)$} \\
		(Tomography)   &                                &                                      &                                       &                                                      \\
		Filtered-QSSFM & \multirow{2}{*}{$\mathcal{O}(\log^2 N)$} & \multirow{2}{*}{Quadratic ($N_t^2$)} & \multirow{2}{*}{$\mathcal{O}(M/\varepsilon^2)$} & \multirow{2}{*}{$\mathcal{O}(N_t^2 M\log^2 N /\varepsilon^2)$} \\
		(Proposed)     &                                &                                      &                                       &                                           
	\end{tabular}
    \end{ruledtabular}
\end{table*}

A rigorous assessment of computational resources requires accounting for the probabilistic nature of quantum measurements and the overhead associated with state preparation in hybrid quantum-classical algorithms.
Unlike classical readout, extracting spectral coefficients via the Hadamard test yields a random variable whose expectation value corresponds to the desired amplitude.
Consequently, the readout process is governed by the sampling complexity required to minimize statistical errors.

Let $M$ denote the number of retained spectral modes and $N_t$ the total number of time steps.
To estimate the spectral coefficients with a precision $\varepsilon$, which is essential for maintaining numerical stability during potential reconstruction, the central limit theorem dictates that the number of measurement shots $N_{\text{shots}}$ must scale as $\mathcal{O}(1/\varepsilon^2)$.
Although quantum amplitude estimation techniques could theoretically improve this scaling to $\mathcal{O}(1/\varepsilon)$, they incur additional circuit depth and are not considered in the standard Hadamard test implementation analyzed here.

Furthermore, the measurement process generally destroys quantum coherence or necessitates the collapse of the wave function.
Therefore, measuring the nonlinear potential at the $t_j$-th time step requires re-preparing the quantum state from the initial state $|\psi(t_0)\rangle$.
This requirement introduces a state-preparation overhead that accumulates throughout the simulation.
At time step $t_j$, the circuit depth is dominated by the linear propagation operators, scaling as $\mathcal{O}(t_j \log^2 N)$.
To advance the simulation from $t=0$ to $t=N_t \tau$, the total runtime $T$ is given by the summation of costs over all steps
\begin{equation}
    T \approx \sum_{t_j=1}^{N_t} M N_{\text{shots}} \mathcal{O}(t_j \log^2 N) \approx \mathcal{O}\left( N_t^2 M \log^2 N/\varepsilon^2 \right).
\end{equation}
This quadratic dependence on the number of time steps, $\mathcal{O}(N_t^2)$, represents the temporal overhead characteristic of hybrid solvers lacking coherent feedback mechanisms.

The complexity comparison is summarized in Tab.~\ref{tab:complexity}.
The classical SSFM scales linearly with $N_t$ and quasilinearly with the system size $N$, yielding $\mathcal{O}(N_t N \log N)$.
In contrast, the filtered-QSSFM exhibits an exponential advantage in spatial scaling but incurs polynomial overheads regarding the desired precision and temporal duration.
Consequently, the proposed framework offers a distinct quantum advantage in the regime of high-dimensional problems ($N \gg 1$) where the exponential spatial speedup outweighs the measurement overheads, provided that the dynamics can be effectively captured by a sparse spectral representation, i.e., $M \ll N$.

\section{\label{sec:4}Results}

In this section, we address three primary aspects of the proposed filtered-QSSFM.
First, we investigate the impact of high-wavenumber truncation on the wave-function evolution.
Second, we analyze the sensitivity of the retention strategy and numerical behavior to variations in the total basis size under a fixed retention setting.
Finally, we evaluate the overall accuracy of the proposed framework relative to the classical SSFM.

\subsection{Numerical setup}
Although the number of retained modes $M$ can in principle be selected arbitrarily, for analytical convenience we parameterize $M$ using the number of retained qubits, defined as $m = \log_2 M$.
In the discrete spectral representation, the initial $M/2$ coefficients correspond to the positive low-frequency components of the spectrum, while the remaining $M/2$ coefficients represent the negative low-frequency components.

We validate the method using four test cases: a 1D soliton, a 2D Gaussian wave packet, 2D snake instability and a moving 2D cylindrical obstacle within a superfluid BEC.
For all cases, periodic boundary conditions are imposed to facilitate the numerical implementation.

\subsection{1D soliton}

To benchmark the performance of the filtered-QSSFM, we consider the evolution of a 1D soliton initialized by the wave function
\begin{equation}
    \psi(x)=\frac{1}{\sqrt{2}}\cosh\left(\frac{x}{\sqrt{2}}\right)\ee^{\ii x}.
\end{equation}
Given that the quantum simulation procedure inherently normalizes the state vector, the coupling coefficient is set to $g = -\|\psi\|_2^2$.
This specification ensures that the physical amplitude of the soliton remains invariant under the normalization imposed by the quantum encoding.

\begin{figure}[t!]
    \centering
    \includegraphics{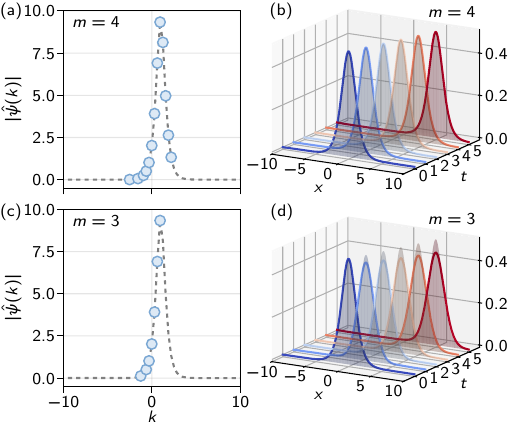}
    \caption{Soliton wave evolution obtained via the filtered-QSSFM with $n=8$ and varying retained qubits $m$.
    The amplitude spectra of the initial state are illustrated for (a) $m=4$ and (c) $m=3$.
    In these panels, blue circles denote the retained modes, while gray dashed lines represent the original modes.
    The corresponding time evolution in physical space over $t \in [0,5]$ is displayed for (b) $m=4$ and (d) $m=3$.
    Solid lines represent the filtered-QSSFM results, whereas the shaded regions indicate the ground-truth solution with the same grid resolution.}
    \label{fig:1D_soliton_m_effect}
\end{figure}

We first investigate the influence of the spectral truncation threshold by fixing the total system size at $n = 8$ qubits and varying the number of retained spectral qubits, denoted by $m$.
We compare two representative settings: $m=3$ and $m=4$.
As illustrated in Figs.~\ref{fig:1D_soliton_m_effect}(a) and \ref{fig:1D_soliton_m_effect}(b), retaining only $m=3$ qubits proves insufficient to resolve the dominant spectral content, leading to noticeable discrepancies between the quantum simulation and the reference evolution.
Moreover, the truncation of high-wavenumber components naturally reduces the discrete norm of the state, which manifests as a global attenuation of the amplitude.
In contrast, Figs.~\ref{fig:1D_soliton_m_effect}(c) and \ref{fig:1D_soliton_m_effect}(d) demonstrate that setting $m=4$ effectively preserves the dominant modes, allowing the quantum solution to reproduce the reference evolution with high fidelity.
Based on these findings, we fix $m=4$ for the subsequent analysis.

\begin{figure*}[t!]
    \centering
    \includegraphics[width=\linewidth]{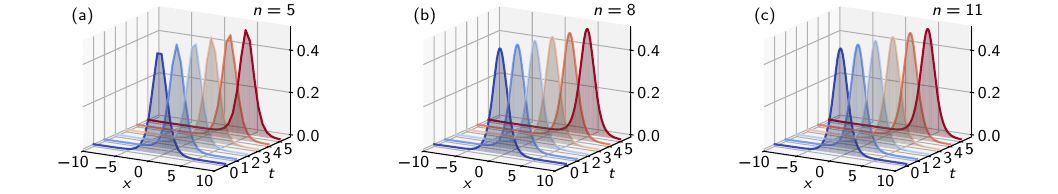}
    \caption{Soliton wave evolution obtained by the filtered-QSSFM with retained qubits $m=4$ and different total qubits: (a) $n=5$, (b) $n=8$, and (c) $n=11$. Solid lines represent the filtered-QSSFM results, whereas the shaded regions indicate the ground-truth solution with the same grid resolution.}
    \label{fig:1D_soliton_n_effect}
\end{figure*}

Next, we verify whether the choice of the total number of qubits $n$ influences the filtering strategy.
Figure~\ref{fig:1D_soliton_n_effect} presents simulation results for $n=5, 8, \text{and } 11$ while keeping the retained window fixed at $m=4$.
For $n=5$ (see Fig.~\ref{fig:1D_soliton_n_effect}(a)), although the overall soliton profile is captured, the wave function lacks smoothness due to limited spatial resolution.
Increasing the system size to $n=8$ (see Fig.~\ref{fig:1D_soliton_n_effect}(b)) yields a smooth and well-resolved wave function.
A further increase to $n=11$ (see Fig.~\ref{fig:1D_soliton_n_effect}(c)) produces no substantial changes in the filtered output.
These results confirm that the filtering strategy depends primarily on $m$, while $n$ dictates the spatial resolution; this observation supports the complexity analysis in Sec.~\ref{subsec:3.5}, which treats $n$ and $m$ as independent parameters.

\begin{figure*}
    \centering
    \includegraphics{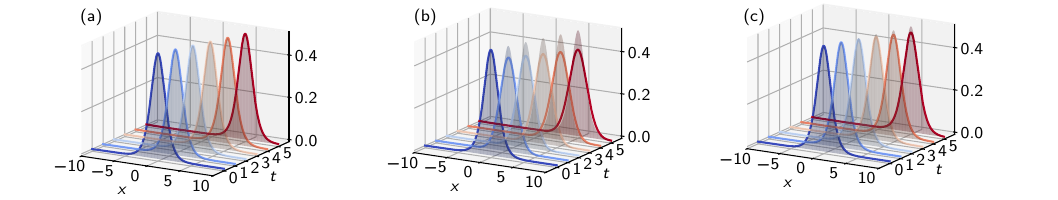}
    \caption{Soliton wave evolution obtained via (a) the QSSFM, (b) the filtered-QSSFM without normalization in the nonlinear step, and (c) the filtered-QSSFM with normalization in the nonlinear step, with $n=8$ and $m=4$. Solid lines represent the filtered-QSSFM results, whereas the shaded regions indicate the ground-truth solution with the same grid resolution.}
    \label{fig:1D_soliton_norm_effect}
\end{figure*}

Finally, we evaluate the necessity of renormalization within the filtered-QSSFM workflow.
Figure~\ref{fig:1D_soliton_norm_effect} compares three simulation regimes using $n=8$ and $m=4$.
The standard QSSFM (see Fig.~\ref{fig:1D_soliton_norm_effect}(a)) serves as the baseline, accurately reproducing both the amplitude and phase of the SSFM reference.
In the filtered-QSSFM without normalization (see Fig.~\ref{fig:1D_soliton_norm_effect}(b)), the nonlinear term is evaluated directly using the unnormalized retained modes.
Consequently, numerical errors accumulate over successive iterations, resulting in pronounced deviations from the exact dynamics.
Figure~\ref{fig:1D_soliton_norm_effect}(c) illustrates the efficacy of the renormalization strategy, where the simulation stability is significantly improved by normalizing the reconstructed wave function prior to evaluating the nonlinear operator.
Although the filtering step inherently truncates high-frequency energy, resulting in a systematic, albeit minor, amplitude decay, the renormalization process numerically compensates for this norm loss.
This compensation ensures that the filtered-QSSFM maintains close agreement with the classical SSFM reference, thereby avoiding the divergence observed in the unnormalized approach.

\begin{figure}[t!]
    \centering
    \includegraphics{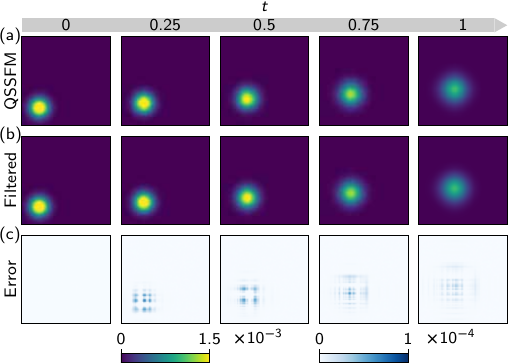}
    \caption{Simulation results for the 2D Gaussian wave evolution at $t=0, 0.25, 0.5, 0.75$, and $1$ using $n_x=n_y=7$ qubits: (a) the QSSFM, (b) the filtered-QSSFM with $m_x=m_y=3$, and (c) the filtering error.}
    \label{fig:2D_Gaussian}
\end{figure}

\begin{figure}[t!]
    \centering
    \includegraphics[width=\linewidth]{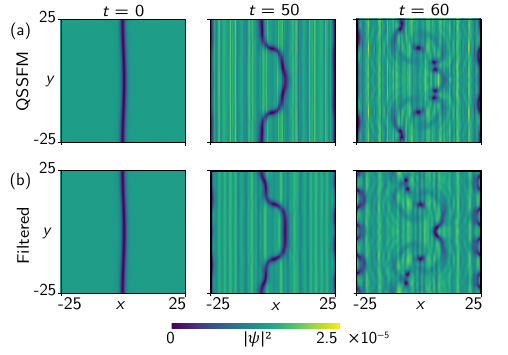}
    \caption{Simulation results for the 2D snake instability evolution at $t=0, 50$, and $60$ using $n_x=n_y=8$ qubits: (a) the QSSFM, (b) the filtered-QSSFM with $m_x=m_y=5$. }
    \label{fig:2D_instability}
\end{figure}

\begin{figure*}[t!]
    \centering
    \includegraphics{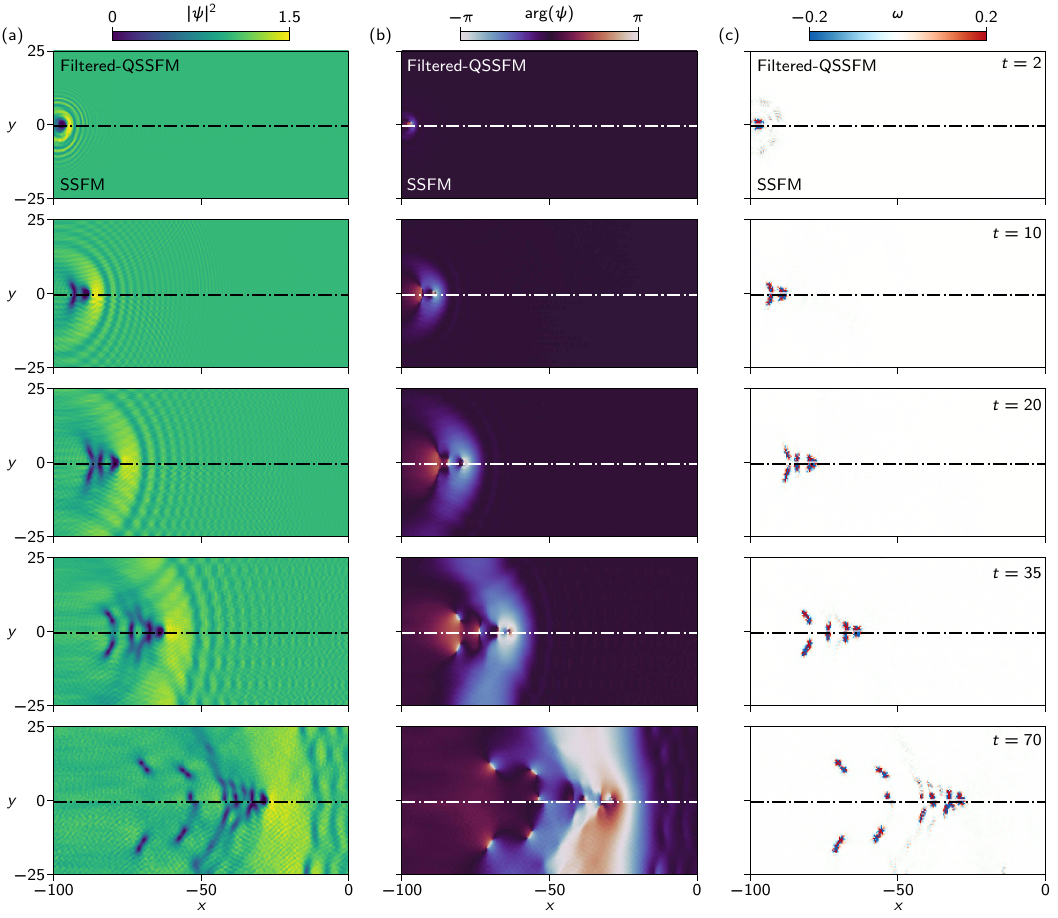}
    \caption{Comparison of simulation results for the 2D wake flow past a cylinder: (a) probability density $|\psi|^2$, (b) phase of the wavefunction $\mathrm{arg}(\psi)$, and (c) vorticity $\omega$ at $t=2, 10, 20, 35$, and $70$.
    Due to the symmetry of the flow about $y=0$, the upper half-plane displays the filtered-QSSFM results computed with $n_x=9$, $n_y=7$, $m_x=7$, and $m_y=5$, while the lower half presents the SSFM results for comparison.}
    \label{fig:2D_cylinder}
\end{figure*}

\subsection{2D Gaussian wave packet}
To demonstrate the applicability of the method beyond one spatial dimension, we consider the evolution of a 2D Gaussian wave packet with initial condition
\begin{equation}
    \psi(x,y) = \frac{1}{2}\ee^{-(x+3)^2 - (y+3)^2} \ee^{\ii (2x + 2y)}.
\end{equation}
In this 2D case, the quantum register comprises $n = 14$ qubits, distributed as $n_x = n_y = 7$ for the respective spatial coordinates.
For the filtering stage, the number of retained qubits is set to $m = 6$, corresponding to $m_x = m_y = 3$ active spectral modes per dimension.
Figure~\ref{fig:2D_Gaussian}(a) displays the time evolution obtained via the standard QSSFM, while Fig.~\ref{fig:2D_Gaussian}(b) presents the results generated by the classically simulated filtered-QSSFM.
The corresponding error distribution is plotted in Fig.~\ref{fig:2D_Gaussian}(c), where the deviation is quantified as
\begin{align}
    \epsilon(x,y) = |\psi_c(x,y)|^2 - |\psi_q(x,y)|^2.
\end{align}
The observed discrepancy is primarily attributed to the reduced effective resolution resulting from the limited spectral bandwidth of the retained qubits.
Nevertheless, the filtered-QSSFM solution successfully preserves the essential qualitative features and maintains a spatial profile consistent with the reference dynamics.

\subsection{2D snake instability}
We investigate the snake instability that emerges from the evolution of a dark soliton solution of the NLSE with parameters $g=1$ and $V=1$.
The computational domain is taken as $x,y\in[-25,25]$, and the time step is $\Delta t=0.01$.
We use $n_x=n_y=8$ qubits for amplitude encoding, and set $m_x=m_y=5$ retained qubits in the filtered-QSSFM.
The initial condition is a perturbed dark soliton
\begin{align}
    \psi(x,y) = \ee^{\ii A\cos{(\xi y)}}\tanh{ \left( x -A\cos{( \xi y)} \right)},
\end{align}
with $L_y=50$, $A=0.03$, and $\xi = 2\pi/L_y$, where the factor $\ee^{\ii A\cos{(\xi y)}}$ introduces the transverse perturbation that seeds the instability.

Figure~\ref{fig:2D_instability} illustrates the results obtained from the QSSFM and filtered-QSSFM at $t=0,50$, and $60$, respectively.
The initial soliton configuration is depicted at $t=0$.
At $t=50$, the onset of instability becomes apparent through the pronounced transverse bending of the dark soliton, which remains spatially continuous.
Subsequently, the soliton undergoes fragmentation at $t=60$, signifying the full development of the snake instability.

However, a discernible discrepancy between the results in Figs.~\ref{fig:2D_instability}(a) and (b) emerges at $t=60$.
This difference is primarily attributed to high-wavenumber modes generated during the evolution of the instability.
These components manifest as high-frequency vertical structures in the snapshots at $t=50$ and $t=60$.
Within the filtered QSSFM framework, such high-frequency contributions are truncated during the mode-filtering stage.
Consequently, the cumulative removal of these modes leads to increasingly prominent deviations between the two methods as the instability develops.

\subsection{2D wake flow past a cylinder}
We further investigate a 2D superfluid BEC interacting with a moving cylindrical obstacle.
Building upon the framework established in Ref.~\cite{connor2025tensor}, we replace the original square barrier with a cylindrical potential to better reproduce the canonical cylinder-flow configuration in fluid dynamics.
Periodic boundary conditions are imposed on the computational domain defined by $x \in [-100,100]$ and $y \in [-25,25]$.
The spatial domain is discretized on a uniform grid utilizing $n_x = 9$ and $n_y = 7$ qubits for the respective directions.
In the spectral filtering stage, the number of retained qubits is set to $m_x = 7$ and $m_y = 5$.
The system is evolved until $t = 100$ with a time step of $\Delta t= 0.01$, and the nonlinearity coefficient is fixed at $g = 1$.
The external potential $V(\vec{x},t)$ represents a circular barrier of radius $r = 1$ and height $V_0 = 10$ translating uniformly in the positive $x$-direction with speed $v_x = 1$.
At $t = 0$, the barrier is centered at $x_0 = -L_x/2 + r $ and $y_0  = 0$.

Figure~\ref{fig:2D_cylinder} presents a comparison between the classical SSFM reference solutions and the results obtained via the filtered-QSSFM.
The density profile $|\psi|^2$, relative phase distribution, and vorticity field are illustrated in Figs.~\ref{fig:2D_cylinder}(a), \ref{fig:2D_cylinder}(b), and \ref{fig:2D_cylinder}(c), respectively.
Distinct vortex shedding is observed, demonstrating that the proposed framework effectively captures the characteristic vortex generation and shedding dynamics inherent to the 2D BEC flow past an obstacle.
These results indicate that the present methodology is extensible to higher spatial dimensions.

Hence, provided high-frequency modes are appropriately truncated, the filtered-QSSFM approach approximates the target wave function and its nonlinear evolution with high fidelity.
Crucially, the number of retained qubits $m$ is effectively decoupled from the total system size $n$: for a prescribed accuracy threshold, the retention strategy remains invariant with respect to $n$.
This decoupling allows $m$ to be treated as an independent parameter in the complexity analysis in Sec.~\ref{subsec:3.5}, thereby offering potential advantages regarding quantum resource utilization and computational overhead.
Consequently, if the wave function is characterized by a compact spectral support, the nonlinear dynamics can be accurately captured by measuring only a reduced subset of quantum states, enabling efficient simulation with substantially fewer measurement requirements.

\section{\label{sec:5}Conclusion}

We propose a hybrid quantum-classical framework for solving the NLSE based on the SSFM.
This approach preserves the essential operator-splitting structure, in which the linear step remains diagonal in the spectral space and the nonlinear term acts as a local phase modulation in position space.
The $\QFT$ is used to switch between the spectral and physical representations.
Consequently, the QSSFM reproduces the key features of the reference SSFM solution, including the wave function profile and phase structure, thereby demonstrating the feasibility of the proposed mapping and offering a unified basis for subsequent algorithmic refinement and resource evaluation.

From the viewpoint of resources and computational complexity, the results indicate that the dominant cost typically arises from the measurement and reconstruction procedures required for state-information acquisition and nonlinear-gate updates, rather than from the linear propagation or Fourier transforms.
If full quantum state tomography is employed to recover the complete state, both the sampling requirement and computational burden grow rapidly.
Consequently, we introduce the strategy of filtering to retain a limited number of modes and reconstructing only the dominant spectral coefficients, thereby reducing the reconstruction scale from the full spatial dimension to the subspace spanned by the retained modes.
We find that treating the number of retained qubits as an independent parameter yields a more scalable cost characterization.
This observation suggests that when the dynamics are dominated by low-frequency structures, one may tune the trade-off between accuracy and reconstruction cost by adjusting the number of retained modes.

Numerical studies further indicate that spectral truncation is a critical factor in the present hybrid framework.
Retaining dominant frequency components improves the agreement between the reconstructed wave function and the classical reference solution regarding the principal structures.
If the nonlinear step evaluates the nonlinear term using unnormalized retained modes, the nonlinear phase update becomes highly sensitive to sampling errors, leading to error accumulation and a pronounced deviation from the reference dynamics.
Normalizing the retained modes stabilizes the estimation of the nonlinear term and substantially reduces cumulative bias, although filtering in the linear step induces appreciable norm loss that results in amplitude decay over time.
Moreover, renormalizing the filtered quantum state at each step numerically compensates for the truncation-induced norm reduction, yielding an evolution closer to the classical SSFM result.

More generally, stable long-time simulation requires simultaneous control of statistical errors in the estimation of the nonlinear term and of the resulting cumulative bias, together with control of norm loss induced by filtering in the linear step.
Statistical errors can be mitigated by improved sampling strategies, higher-precision estimation, or by employing more robust schemes for nonlinear phase updates.
Norm loss, however, necessitates an explicit trade-off between physical fidelity and numerical stability, for example via stepwise re-normalization or norm correction to limit its impact on amplitude evolution.
Overall, these observations indicate clear directions for future algorithm design: reduce the sensitivity of nonlinear updates to reconstruction errors while preserving the effective contributions of key conserved quantities under spectral reduction.

Looking forward, the proposed framework can be naturally extended to higher-dimensional NLSEs.
Moreover, given the broad relevance of nonlinear wave phenomena, the QSSFM structure may serve as a practical template for hybrid quantum-classical simulation of other nonlinear wave systems.
More broadly, achieving stable updates of nonlinear feedback terms under controlled reconstruction costs remains a central challenge for realizing high-accuracy, long-time evolution in hybrid frameworks.
The numerical implementation, error-source analysis, and complexity characterization presented here provide a methodological foundation for further advances.

\begin{acknowledgments}
ZM acknowledges support from NSFC the Excellence Research Group Program for multiscale problems in nonlinear mechanics (Grant No.~12588201).
GH acknowledges support from NSFC (Grant No. 12332016).
\end{acknowledgments}

\bibliography{main.bib}% Produces the bibliography via BibTeX.

\end{document}